\documentclass{JHEP}
\usepackage{amsmath}
\usepackage{amssymb,amsfonts}
\usepackage{epsfig}

\newcommand{\Zint}{\mathbb{Z}}
\newcommand{\Real}{\mathbb{R}}
\newcommand{\Comp}{\mathbb{C}}

\newcommand{\be}{\begin{equation}}
\newcommand{\ee}{\end{equation}}
\newcommand{\bea}{\begin{eqnarray}}
\newcommand{\eea}{\end{eqnarray}}
\newcommand{\CP}{\mathbb{C}\mathbb{P}}

\newcommand{\C}{{\cal C}}
\newcommand{\D}{{\cal D}}

\newcommand{\G}{{\cal G}}

\newcommand{\M}{{\cal M}}
\newcommand{\N}{{\cal N}}
\newcommand{\Sy}{{\cal S}}
\newcommand{\Ka}{K{\"a}hler }

\newcommand{\ub}{\overline{1}}

\newcommand{\uh}{\widehat{1}}
\newcommand{\ch}{\widehat{5}}
\newcommand{\sh}{\widehat{7}}
\newcommand{\uf}{1^{^{\!\!\!\!\!\!\frown}}}
\newcommand{\us}{1_{_{_{\!\!\!\!\!\smile}}}}
\newcommand{\xh}{\hat{x}}
\newcommand{\xb}{\bar{x}}
\newcommand{\xhb}{\bar{\hat{x}}}
\newcommand{\xhh}{\hat{\hat{x}}}
\newcommand{\yh}{\hat{y}}
\newcommand{\yb}{\bar{y}}

\newcommand{\zb}{\bar{z}}
\newcommand{\yhb}{\bar{\hat{y}}}
\newcommand{\yhh}{\hat{\hat{y}}}
\newcommand{\fb}{\bar{f}}
\newcommand{\eb}{\bar{e}}
\newcommand{\Co}{\mathbb{C}}
\newcommand{\I}{{\cal I}}
\renewcommand{\uf}{1^{^{^{\!\!\!\!\!\frown}}}} 
\renewcommand{\us}{1_{_{_{\!\!\!\!\!\smile}}}}
\newcommand{\ufp}{1^{^{\!\!\!\!\!\frown}}} 
\newcommand{\usp}{1_{_{_{\!\!\!\!\!\smile}}}}

\newcommand{\pamatrix}[1]{\begin{pmatrix} #1 \end{pmatrix}}

\title{Rolling Among $G_2$ Vacua}

\author{Herv{\'e} Partouche\\
Centre de Physique Th{\'e}orique, Ecole Polytechnique,\\
F-91128 Palaiseau CEDEX, France\\
E-mail: \email{Herve.Partouche@cpht.polytechnique.fr}}

\author{Boris Pioline\footnote{On
leave of absence from LPTHE, Universit{\'e} Pierre et Marie Curie,
PARIS VI and Universit{\'e} Denis Diderot, PARIS VII, Bo{\^\i}te
126, Tour 16, 1$^{\it er}$ {\'e}tage, 4 place Jussieu, F-75252
Paris CEDEX 05, FRANCE}\\
Jefferson Physical Laboratory, Harvard University\\
Cambridge, MA 02138, USA\\
E-mail: \email{pioline@physics.harvard.edu}}

\abstract{We consider topology-changing transitions between 7-manifolds
of holonomy $G_2$ constructed as a quotient of $CY\times S^1$ 
by an antiholomorphic involution. We classify involutions for
Complete Intersection CY threefolds, focussing primarily
on cases without fixed points.  The ordinary conifold transition 
between CY threefolds descends to a transition between $G_2$ manifolds, 
corresponding
in the $\N=1$ effective theory incorporating the light black hole states
either to a change of branch in the scalar potential or to a Higgs
mechanism. A simple example of conifold transition with
a fixed nodal point is also discussed.
As a spin-off, we obtain examples of $G_2$ manifolds
with the same value for the sum of Betti numbers $b_2+b_3$,
and hence potential candidates for mirror manifolds.}

\preprint{hep-th/0011130\\CPHT-S091.1100\\HUTP-00/A044\\LPTHE-00-42}

\begin{document}

\section{Introduction}
One of the important discoveries of the past few years in string theory
is that a vast number of consistent string models are not disconnected,
but actually different vacua related by smooth deformations or more
radical phase transitions. This view has emerged 
from a better understanding of the non-perturbative spectrum of 
string theories and of their non-perturbative dualities. 
In particular, for type II string theories with $\N=2$ supersymmetry
in four dimensions, it has been found that a wide class of Calabi-Yau (CY) 
compactifications
were related by smooth topology-changing ``conifold'' 
transitions \cite{Candelas:1990ug},
whereby a two-cycle shrinks to zero size and reappears as a
three-cycle (see also \cite{Green:1988wa, Green:1988bp,Candelas:1989di}).
While this transition was known at the mathematical 
level \cite{lefschetz}, string
theory gives a smooth representation of this transition by providing
the low-energy degrees of freedom that resolve the singularity on
the conformal field theory moduli space: namely, the Ramond-Ramond 
charged black holes that become massless 
at the singularity \cite{Strominger:1995cz,Greene:1995hu}. 
The purpose of this work
is to study such topology-changing transitions in a $\N=1$ setting.

$\N=1$ vacua in four dimensions can be constructed in many ways.
The heterotic string compactified on a CY threefold,
possibly in the presence of 5-brane sources, has been intensively
studied but is complicated by the fact that a superpotential
may be generated already at tree-level by 
worldsheet instantons \cite{Dine:1986zy,Witten:1996bn}.
Type I strings or orientifold of type II on CY threefolds
are another option, and conifold transitions were in particular considered
in type I' string theory in \cite{Vafa:1996gm}. 
This can also be reformulated geometrically as 
compactifications of F-theory on a CY fourfold \cite{Donagi:1996yf},
or by considering space-filling branes wrapped on 
supersymmetric cycles in type II
compactifications on CY threefolds \cite{Brunner:2000jq,
Kachru:2000ih,Kachru:2000an}.
In this work, we consider another geometric realization,
namely M-theory compactified on
7-manifolds of exceptional holonomy $G_2$. Using
the invariance under Peccei-Quinn-type symmetries,
one can argue that in this case the superpotential
may arise only at the non-perturbative level from membrane
instantons \cite{Becker:1995kb,Harvey:1999as}. Joyce has proposed a relatively
simple construction of $G_2$ manifolds \cite{Joyce1,Joyce2}, as a quotient
\begin{equation}
\label{g2}
\G= ( \C \times S^1 )/\Zint_2
\end{equation}
of the product of a CY threefold $\C$ and
a circle $S^1$ by an involution $\sigma=w\I$ acting as an 
inversion $\I$: $x^{10}\rightarrow -x^{10}$ 
on $S^1$ and as an antiholomorphic involution $w$ on $\C$
such that $w^*(J)=-J$ and
$w^*(\Omega)=e^{i\theta} \bar{\Omega}$, where $J$ 
and $\Omega$ are the \Ka form and holomorphic three-form, 
while $\theta$ is a real constant. The main goal of this paper
is to study topology-changing transitions between
$G_2$ manifolds of the form \eqref{g2} resulting
from conifold transitions in $\C$. We shall focus
on the simplest (Abelian) type of transition, but our discussion
could easily be generalized to more complicated non-Abelian
transitions \cite{Bershadsky:1996sp,Klemm:1996kv,
Katz:1996ht,Berglund:1997uy}. Compactifications on singular $G_2$ manifolds
have also been considered from the point of view of geometric
engineering in\footnote{As this work was finalized, a preprint
appeared \cite{Acharya} which discusses non-Abelian singularities
in local models of $G_2$ manifolds.}  \cite{Acharya:1998pm}. 
We shall also disregard the superpotential that might be generated
on either side of the transition by instantons.
Such a superpotential may lift part or all of the
branches of the moduli space on either side, and in particular
drive the theory to the conifold point. At any rate, it 
does not prevent a continuous transition between 
the remaining vacua in {\it configuration} space,
which is a necessary condition for the existence of tunneling processes
between $\N=1$ vacua.

Besides extremal transitions, CY manifolds can also
be related by mirror symmetry. This is not a continuous transition between
different CY's proper, but rather a smooth cross-over between different
geometric descriptions of a same CFT in different regime of moduli space.
Mirror symmetry has also been conjectured to hold in the context
of $G_2$ manifolds \cite{Shatashvili:1994zw, Acharya:1998rh}, 
although much less is known, for lack of
a precise understanding of the CFT. Although this is somewhat 
peripheral to the focus of the present work, we shall 
also exhibit various examples of $G_2$ manifolds, whose Betti numbers satisfy
$b_2+b_3=$ constant. This relation is a necessary condition for $G_2$ 
manifolds to be mirror \cite{Shatashvili:1994zw}, 
and it would be interesting to
investigate whether the CFT's underlying these examples are indeed
equivalent.

The plan of this work is as follows. We start in Section 2
by recalling some background about $G_2$ manifolds,
discuss Joyce's construction and classify possible
antiholomorphic involutions. Section 3 is a short
review of conifold transitions in Complete Intersection
Calabi-Yau (CICY) manifolds on which we focus in the
following. In Section 4, we discuss transitions
between $G_2$ manifolds triggered by a conifold transition
in the underlying CICY, both from the mathematical and
physical point of view. For the latter, we show
that conifold transitions correspond to either 
a change of scalar branch in the $\N=1$ moduli space, or to a
standard Higgs effect. In section 5, we briefly discuss
the case where the involution has a non-empty fixed point
set, and in particular when it contains the nodal points of the
conifold.

\section{$G_2$ manifolds as Calabi-Yau orbifolds \label{invol}}

\subsection{General facts about $G_2$ manifolds}
Seven-manifolds of exceptional holonomy $G_2\subset SO(7)$ 
have one covariantly constant spinor $\theta$, as apparent from
the branching rule ${\bf 8}={\bf 7}\oplus {\bf 1}$ of the
spinor representation of $SO(7)$.
Equivalently, there is one covariantly constant three-form
$\phi$, closed and co-closed \cite{Gibbons:1990er}. Locally, one may choose an orthogonal
frame $e_i$ in which
\begin{equation}
\phi=e_{127}+e_{136}+e_{145}+e_{235}+e_{426}+e_{347}+e_{567}\ ,\quad
e_{ijk}=e_i\wedge e_j \wedge e_k\ ,
\end{equation}
where we recognize on the r.h.s. the structure constants of
the unit octonions. 
Compact $G_2$ manifolds provide $\N=1$ supersymmetric backgrounds for classical
eleven-dimensional supergravity. The massless
spectrum follows from simple homological considerations
\cite{Papadopoulos:1995da}. 
The deformations of the three-form $\phi$ yield $b_3$
real moduli, which combine with the flux of the 
3-form $C$ and the modes of the gravitino
into $b_3$ $\N=1$ chiral multiplets. In addition,
the reduction of the three-form on the $b_2$ 2-cycles
yields $b_2$ gauge fields, which together with 
the reduction of the gravitino make up $b_2$ $\N=1$
vector multiplets. The \Ka potential and gauge kinetic
term are simply obtained from the volume $V$ of the
manifold and intersection matrix respectively,
while the superpotential vanishes in the 
classical supergravity approximation. By the usual
arguments of holomorphy and Peccei-Quinn symmetry, it remains
zero to all orders in $1/V$, but may be generated
by membrane instantons \cite{Harvey:1999as}.

\subsection{Joyce's construction of $G_2$ manifold}
\label{cons}
The first examples of compact $G_2$ manifolds have
been constructed by orbifold constructions $T^7/\Gamma$,
where $\Gamma$ is a discrete group commuting with
a $G_2$ subgroup of $SO(7)$ only \cite{Joyce1, Joyce2}. The fixed points
singularities can be resolved by appropriately gluing in  
Eguchi-Hanson spaces with $SU(2)$ holonomy, so that the total holonomy
lies in all of $G_2$. This construction can be repeated
by orbifolding other compact 7-manifolds with reduced holonomy
such as $K_3\times T^3$ or, more generally, $\C\times S^1$ for a CY
threefold $\C$. One thus considers quotients
\begin{equation}
\label{g22}
\G= ( \C \times S^1 )/\sigma
\end{equation}
where $\sigma=w\I$ is an involution 
acting as $\I$: $x\to -x$ on $S^1$ and antiholomorphically on $\C$.
$\sigma$ must in addition be an isometry, so that
$w^*(J)=-J$ and $w^*(\Omega)=e^{i\theta}\bar\Omega$.
The closed and co-closed four-form
\begin{equation}
\phi=  J \wedge dx + \Re(e^{-i\theta/2} \Omega)
\end{equation}
is invariant under $\sigma$, and provides the quotient $(\C\times
S^1)/\Zint_2$ with a $G_2$ structure.

In general, the involution $w$ may have a
non-empty fixed point set $\Sigma$ on $\C$, which is then
a compact special Lagrangian 3-cycle \cite{har-law}.
In that case, one must in addition resolve the singularity of the quotient,
by gluing in an Eguchi-Hanson space in the space transverse
to $\Sigma$. It appears necessary to have $b_1(\Sigma)>0$ in order
for a resolution to exist. This is not surprising, since this is
also the number of deformations of the special Lagrangian 3-cycle
$\Sigma$ \cite{Mc Lean}. In the following, we concentrate
on orbifolds without fixed points, but we will return to the
problem of fixed points in Section 5.

The Betti numbers of the manifold in \eqref{g22} can 
be counted as follows.
Let us denote $h_{11}$, $h_{12}$ the Hodge numbers 
of the CY and $h_{11}^\pm$ the
number of even (odd) two-forms. The number of invariant two-forms on
$(\C \times S^1)/w\I$ is then simply $h_{11}^+$, while three-forms are
obtained by wedging the $h_{11}^-$ odd two-forms 
on $\C$ with $dx$. Moreover, the real parts of the three-form
in $H^{1,2}(\C)\oplus H^{2,1}(\C)$ and
$H^{0,3}(\C) \oplus H^{3,0}(\C)$ are also invariant. 
The untwisted Betti numbers of $\G$ are therefore 
\begin{equation}  
b_2=h_{11}^+ \quad \mbox{and}\quad b_3=1+h_{11}^-+h_{12}\ .
\label{sp}
\end{equation}
In the non-freely acting case, one has
to add the contribution of the desingularization of
the fixed points, as we discuss in Section 5.

For simplicity, we will restrict our attention to $G_2$ manifolds
constructed out of Complete Intersection CY (CICY) threefolds
\cite{Green:1987ck, Candelas:1988kf}, {\em i.e.} defined  
by a set of homogeneous equations in a product of projective spaces
${\cal P}=\prod_{k=1}^{r} \CP^{n_k}$. The full
moduli space to which such a manifold belongs is specified by a
configuration matrix of the form 
\begin{equation}
\label{cm}
\left[
\begin{array}{l||l}
V&D
\end{array}
\right]^{h_{11}, h_{12}}_{\chi}\; ,
\end{equation}
where $V$ is a column $r$-vector, whose entries are the dimensions of
the $r$ embedding $\CP^{n_k}$ factors, and $D$ is a $r\times h$
matrix, where $h=\sum_{k=1}^{r} n_k-3$ is the number of homogeneous
polynomial constraints on the projective coordinates. The entry
$D_{k,l}$ is the degree of the $l$-th constraint in the
homogeneous coordinates of $\CP^{n_k}$. $\chi=2(h_{11}-h_{12})$ is the
Euler characteristic, which is easily computed from the matrix
\eqref{cm} \cite{Green:1989cr, Candelas:1988kf}.
In order for the manifold to be Ricci flat, we
must have $\sum_{l=1}^{h} D_{kl}=n_k+1$, for all $k=1,...,r$. 
A particular set of antiholomorphic involutions is then obtained
by restricting antiholomorphic involutions of the projective space 
${\cal P}$.
Requiring this involution
to be an isometry of the CY will restrict the allowed coefficients
of the homogeneous equations.

\subsection{Antiholomorphic involutions of projective spaces\label{invproj}}
Let us start by classifying the antiholomorphic involutions $w$
of a single projective space $\CP^n$. We represent them by
the matrix $M$ such that $z_i \to M_{ij} \bar z_j$. $M$
and $\rho M$ define the same involution of $\CP^n$ for any 
$\rho\in \Comp^*$,
so that we may choose $\det M=1$. These
involutions have to be classified up to a holomorphic change of basis
$z_i \to U_{ij} z_j$, which amounts to $M\to U^{-1} M U^*$,
where $U\in Gl(n+1,\Comp)$. Requiring
$w$ to square to 1 projectively imposes $M M^*=\lambda I$.
Taking the trace and determinant of this equation, we see that
$\lambda$ is real and $\lambda^{n+1}=1$. For $n$ even, this
forces $\lambda=1$, while $n$ odd allows for the two possibilities
$\lambda=\pm 1$. Requiring furthermore $w$ to be an isometry
({\em i.e.} preserving the Fubini-Study metric of $\CP^n$)
imposes $M M^\dagger = \mu I$, where $\mu$ is fixed to $1$ by 
the previous choices. Combining this equation with $M M^*=\lambda$
implies that $M$ is symmetric for $\lambda=1$ and antisymmetric
for $\lambda=-1$. In either cases, the real and imaginary parts
of $M$ commute again due to $M M^*=\lambda I$. They can therefore
be simultaneously brought into a diagonal form for $\lambda=1$
or antisymmetric diagonal form for $\lambda=-1$ by a real orthogonal
rotation, hence an allowed holomorphic change of basis. 
Finally, the phase of the coefficients can be reabsorbed by
an holomorphic change of basis. Altogether, we have thus
found two distinct antiholomorphic involutions,
\begin{equation}
\begin{array}{lll}
A:&& (z_1,z_2,\dots,z_{n+1}) \to (\bar z_1,\bar z_2,\dots,\bar{z}_{n+1})\\
B:&& (z_1,z_2,\dots,z_{n},z_{n+1}) \to (-\bar z_2, \bar
z_1,\dots,-\bar z_{n+1}, \bar z_n)\; .
\end{array}
\end{equation}
The two cases correspond to different values of $\lambda$ and 
cannot be combined in the same projective factor $\CP^n$
without spoiling the involution property. 
In particular, the involution $B$ is defined for $n$ odd only, 
and exchanges the projective coordinates by pairs. These
two involutions have very different properties, since
$A$ admits a fixed set $\left\{ z_i=\bar z_i \right\}$, while
$B$ acts freely. For $n=1$, these are the reflection along
the equator and the antipodal map of the sphere $S^2$, respectively.

Finally, we may consider antiholomorphic 
involutions that mix different factors in $\prod_{k=1}^r \CP^{n_k}$.
The involution must commute with the projective actions,
so that the only possibility is to exchange two identical
projective factors,
\begin{equation}
C: \quad (\left\{y_i\right\};\left\{z_i\right\})
\to (\left\{\bar{z}_i\right\};\left\{\bar{y}_i\right\})\; .
\end{equation}
This involution has a fixed point set $\left\{ y_i=\bar z_i \right\}$,
which is the diagonal in $\CP^n \times \CP^n$.

\subsection{Antiholomorphic involutions of CICY\label{nota}}
Having constructed antiholomorphic involutions of projective 
spaces $A,B,C$, we can now combine them to construct $G_2$
manifolds from CICY 3-folds as in \eqref{g22}.
We shall denote the resulting manifolds by the configuration
matrix of the underlying CY, denoting by  $\overline{n}$, $\widehat{n}$,
and  $\begin{array}{c} n^{^{\!\!\!\!\!\!\frown}}\\ 
n_{_{_{\!\!\!\!\!\!\smile}}} \end{array}$ the projective spaces in
the first column on which the involution acts by $A,B$ or $C$,
respectively. For example, the configuration matrix
\begin{equation}
\left[  \begin{array}{c||cccccc} \sh &1&1&1&1&2&2 \\ \uf &1&1&0&0&0&0
    \\ \us &0&0&1&1&0&0 \end{array} 
\right]^{1,54} ,\end{equation}
denotes the family of $G_2$ manifolds constructed from the
CICY with the same configuration matrix, by acting with the
involution $B$ on $\CP^7$ and $C$ on $\CP^1\times \CP^1$
(this example will be treated in detail in Section 4.1.3).
The superscripts indicate the Betti numbers $b_2$ and $b_3$ 
counting the two-and three-cycles invariant under the 
involution, respectively. When the  
involution has fixed points, it is necessary to add the contribution
of the singularities after resolution in order to obtain the
correct topological invariants.
 
In order for the projective space involution $w$ to restrict
to the CY 3-fold, reality conditions must be enforced on the
coefficients of the homogeneous equations. This generically
halves the number of allowed complex deformations of the
CY. In some cases however, there is simply no choice of
coefficients which are preserved by the involution
An example of this is the matrix
\begin{equation}
\left[  \begin{array}{c||ccccc} \ch &3&1&1&1 \\ \ub &1&1&0&0 \\ \uh
    &0&0&1&1   \end{array} 
\right]^{0,73}
\end{equation}
for which it is easy to convince oneself that there is no choice
of equation of bidegree $(3,1)$ in the coordinates of $\CP^5\times \CP^1$ 
compatible with the involution. It is thus important in practice to check
that the projective space involution is compatible with the
CICY configuration matrix.

A second important remark is that the topology of the $G_2$ manifold
is not fully specified by the configuration matrix.
Instead, the moduli space of $G_2$ manifolds associated to a given 
configuration matrix in general splits into several disconnected
components whose Betti numbers differ from each other by
the contribution of fixed points. This is because the locus of
fixed points under a given involution $w$ undergoes 
transitions in real codimension 1, whereby real roots collide
and become imaginary in complex conjugate pairs. 
In the following, we shall
consider cases where the choice of involution ensures that there is
no fixed point throughout the CY moduli space,
so that one obtains only one $G_2$ manifold moduli space.
We postpone to Section 5 the discussion of the more challenging
case where this assumption is not valid.

\section{A review of conifold transition between CICY's\label{CYcase}}

The aim of this section is to review in some detail some
of the mathematical aspects of conifold singularities and transitions between
complete intersection CY's \cite{Candelas:1990ug}. 
A similar approach will then be
taken in the next Section in the $\N=1$ case.

\subsection{An example\label{example}}
Let us consider a CY manifold $\C_1$ chosen at a generic point of the
moduli space $\M^{(1)}$ associated to the configuration matrix  
\begin{equation}
\M^{(1)} =  \left[  \begin{array}{c||ccccc} 7 &1&1&2&2&2 \\ 1 &1&1&0&0&0   \end{array}
\right]^{2,58}_{-112} \cdot
\label{M1}
\end{equation}
Define $x_i$ $(i=1,...,8)$ and $y_j$ $(i=1,2)$ the projective
coordinates in $\CP^{7}$ and $\CP^1$, respectively. The 
defining equations of the manifold take then the form
\begin{equation}
(\Sy_1)\left\{ \begin{array}{l}
f_1(x,y) := P_{11}(x)y_1+P_{12}(x)y_2 =0 \\
f_2(x,y) := P_{21}(x)y_1+P_{22}(x)y_2 =0 \\
e_{2}(x)= e_{3}(x) = e_{4}(x) =0 \; ,
\end{array} \right.
\label{eq}
\end{equation}
where the $P_{k,l}$ $(k,l=1,2)$ and $e_n$ $(n=2,3,4)$ are homogeneous
polynomials in $x_i$'s of degree one and two, respectively. For generic
coefficients, these equations are transverse,
which means that $f_i=e_n=0$ together with $df_1\wedge df_2 \wedge de_2\wedge
de_3\wedge de_4 =0$ has no solution. 
Changing the complex coefficients appearing in the defining
polynomials amounts to changing the complex structure of the
manifold\footnote{Actually, there is not in general a one-to-one
correspondence between the independent polynomial deformations and the complex
structure moduli. See \cite{Green:1987rw, 
Green:1989cr} for details.}.
The K{\"a}hler moduli on the other hand correspond to the volumes $v_7$
and $v_1$ of $\CP^7$ and $\CP^1$, respectively (together with the fluxes
of the three-form $C$ on the three-cycles dual to $J_{1,7}\wedge dx^{10}$).

Since $y_j$ $(j=1,2)$ are projective coordinates, in order
to have non-vanishing solutions to $f_1=f_2=0$ in (\ref{eq}), the matrix of
coefficients $P_{k,l}$ must have vanishing determinant
\be 
e_{1}^\sharp (x) := P_{11}(x)P_{22}(x)-P_{21}(x)P_{12}(x)=0 \; .
\label{det}
\end{equation}
We may thus
dispose of the $\CP^1$ coordinates $(y_1,y_2)$ altogether,
and rewrite the system as
\begin{equation}
(\Sy_0^\sharp)\left\{ \begin{array}{lll}
e_1^\sharp(x) &=& 0 \\
e_{2}(x)&=& e_{3}(x) = e_{4}(x) =0 \; ,
\end{array} \right.
\label{eqsing}
\end{equation}
where the variables $y_{1,2}$ have been ``integrated out''.
This amounts to having shrunk the  $\CP^1$ parameterized
by $y_{1,2}$ to zero size (in particular, the K{\"a}hler 
class of the $\CP^1$ has now disappeared).
We are then left with a variety $(\C_0^\sharp)$ in 
\begin{equation}
\M^{(0)}=\left[
  \begin{array}{c||cccc} 7 &2&2&2&2  \end{array}
\right]^{1,65}_{-128} \cdot 
\label{C0}
\end{equation}
This operation is called a {\em determinantal contraction}, 
and is denoted by
\begin{equation}
\left[  \begin{array}{c||cccc} 7 &2&2&2&2  \end{array}
\right]^{1,65}_{-128} 
\leftarrow
\left[  \begin{array}{c||ccccc} 7 &1&1&2&2&2 \\ 1 &1&1&0&0&0   \end{array}
\right]^{2,58}_{-112} \; ,
\label{split1}
\end{equation}
(the reversed operation being called {\em determinantal splitting}).
Conversely, at a generic point on \eqref{C0}, the matrix
\begin{equation}
\pamatrix{P_{11}(x) & P_{12}(x) \cr P_{21}(x) & P_{22}(x)}
\end{equation}
has rank one: It therefore determines a unique projective solution
of \eqref{eq}, which means that a point in $\C_0^\sharp$ corresponds
to a point in $\C_1$. However, when all $P_{k,l}$ vanish,  the space
\eqref{eqsing} is singular, and there is a full $\CP^1$-worth of
$(y_1,y_2)$ satisfying \eqref{eq}. Since
$P_{k,l}(x)=0$ $(k,l=1,2)$ and $e_n(x)=0$ $(n=2,3,4)$ 
give us 7 conditions for 7 inhomogeneous coordinates in $\CP^7$,
this happens at isolated points on $\C_0^\sharp$ known as
nodal points. A simple counting shows that there are 8 of them.
The manifold $\C_1$ therefore gives a resolution of the
singular manifold $\C_0^\sharp$, where $\CP^1$'s are glued at
each of the nodes.

There is actually another way to desingularize
$\C_0^\sharp$, that is to change the coefficients of the
degree two polynomial $e_1^\sharp$. The deformed space is then defined by
\begin{equation}
(\Sy_0)\left\{ \begin{array}{lll}
e_1(x) & :=& e_1^\sharp(x) - t\varepsilon^2(x) = 0 \\
e_{2}(x)&=& e_{3}(x) = e_{4}(x) =0 \; ,
\end{array} \right.
\label{eqreg}
\end{equation}
where $t$ is some sufficiently small but not zero real number and
$\varepsilon^2(x)$ is any homoneneous polynomial of degree 2 chosen
such that it is non zero at any of the 8 singular points of
$\C_0^\sharp$. We have thus {\em deformed} $\C_0^\sharp$
to a smooth manifold $\C_0$ in $\M^{(0)}$. 

The resolution and deformation described above in fact correspond
to the two ways of desingularizing the local neighborhood 
of each node, which is homeomorphic to a real cone over 
$S^2\times S^3$. In $\C_1$ the apex of each cone is 
blown-up into a sphere $S^2$, while in $\C_0$ the apex
is blown up into a sphere $S^3$.
The transition between the
two is known as a conifold transition. The change in the Euler
characteristic 
accross the transition between two CY's $\C$ and $\C'$ is then in
general simply understood: 
Since $\chi(S^2)=2$, while $\chi(S^3)=0$, we have 
\begin{equation}
{1\over 2}[\chi(\C)-\chi(\C')]=N\; ,   
\label{euler}
\end{equation}
where $N$ counts the number of nodes at the transition.
The change in the Hodge number is more difficult to compute,
and will be explained in Section \ref{bhres}.

\subsection{The web of complete intersection CY's\label{CYweb}}

The determinantal splitting illustrated in the example of the previous
section can now be repeated successively for each of the
degree two equations $e_{2,3,4}$. One then obtains a sequence of transitions
$\M^{(0)} \leftarrow \M^{(1)} \leftarrow \cdots \leftarrow \M^{(4)}$ 
connecting various moduli spaces carracterized by different Hodge
numbers:
\begin{eqnarray}
&
\left[  \begin{array}{c||cccc} 7 &2&2&2&2  \end{array}
\right]^{1,65}_{-128} 
\leftarrow
\left[  \begin{array}{c||ccccc} 7 &1&1&2&2&2 \\ 1 &1&1&0&0&0   \end{array}
\right]^{2,58}_{-112}
\leftarrow
\left[  \begin{array}{c||cccccc} 7 &1&1&1&1&2&2 \\ 1 &1&1&0&0&0&0 \\  1 &0&0&1&1&0&0  \end{array}
\right]^{3,51}_{-96}
\nonumber
\\
\label{seq}
\\
&
\leftarrow
\left[  \begin{array}{c||ccccccc} 7 &1&1&1&1&1&1&2 \\ 1 &1&1&0&0&0&0&0
    \\  1 &0&0&1&1&0&0&0 \\  1 &0&0&0&0&1&1&0 \end{array}
\right]^{4,44}_{-80}
\leftarrow
\left[  \begin{array}{c||cccccccc} 7 &1&1&1&1&1&1&1&1 \\ 1 &1&1&0&0&0&0&0&0
    \\  1 &0&0&1&1&0&0&0&0 \\  1 &0&0&0&0&1&1&0&0 \\ 1 &0&0&0&0&0&0&1&1 \end{array}
\right]^{5,37}_{-64}
\rightarrow
\left[  \begin{array}{c||c}
                1&2 \\
                1&2 \\
                1&2 \\
                1&2 
                \end{array}     \right]^{4,68}_{-128} \cdot
\nonumber
\end{eqnarray} 
In this sequence, the last moduli space we denote $\M^{(1111)}$ was
obtained by sending the volume $v_7$ of 
$\CP^7$ to zero. It is again obtained by determinantal contraction as follows:
write the system of 8 equations associated to  $\M^{(4)}$. Since these
equations are linear in $x_i$ $(i=1,...,8)$ and that projective
coordinates in $\CP^7$ cannot vanish simultaneously, 
the $8\times 8$ determinant of coefficients of the $x_i$'s must be
zero. Denoting by $y_j$, $z_j$,
$t_j$ and $u_j$ $(j=1,2)$ the 
projective coordinates of the 4 $\CP^1$'s used in the definition of
$\M^{(4)}$, 
this determinant $\D^\sharp$ is an homogeneous polynomial of degree 2 in each
of the variables $y$, $z$, $t$ and $u$. 
When the volume of $\CP^7$ is shrunk to zero,
we may integrate out the variables $x_i$ and replace
the 8 equations by the single quadratic equation
\begin{equation}
\D^\sharp(y,z,t,u)=0\ .
\label{D}
\end{equation}
This defines as in the previous section a singular variety in
$\M^{(1111)}$ that can be deformed to a generic smooth manifold.

Actually, the moduli space $\M^{(1111)}$ plays a central role in the
construction of the web of CY's since {\em any} CICY moduli space can be
related to it by performing a finite number of determinantal splittings
and contractions. We conclude this section by giving another
example of sequence we shall consider later, 
originally considered in \cite{Candelas:1990ug}. 
This sequence looks very similar to the one above but
will yield different patterns for its $G_2$ descendants:
\begin{eqnarray}
&
\left[  \begin{array}{c||cc} 5 &4&2  \end{array}
\right]^{1,89}_{-176} 
\leftarrow
\left[  \begin{array}{c||ccc} 5 &4&1&1 \\ 1 &0&1&1   \end{array}
\right]^{2,86}_{-168}
\leftarrow
\left[  \begin{array}{c||cccc} 5 &3&1&1&1 \\ 1 &1&1&0&0 \\  1 &0&0&1&1 \end{array}
\right]^{3,69}_{-132}
\nonumber
\\
\label{seq'}
\\
&
\leftarrow
\left[  \begin{array}{c||ccccc} 5 &2&1&1&1&1 \\ 1 &1&1&0&0&0 
    \\  1 &0&1&1&0&0 \\  1 &0&0&0&1&1 \end{array}
\right]^{4,46}_{-84}
\leftarrow
\left[  \begin{array}{c||cccccc} 5 &1&1&1&1&1&1  \\ 1 &1&1&0&0&0&0 
    \\  1 &0&1&1&0&0&0  \\  1 &0&0&1&1&0&0 \\ 1 &0&0&0&0&1&1 \end{array}
\right]^{5,37}_{-64}
\rightarrow
\left[  \begin{array}{c||c}
                1&2 \\
                1&2 \\
                1&2 \\
                1&2 
                \end{array}     \right]^{4,68}_{-128} .
\nonumber
\end{eqnarray} 

\subsection{Black hole resolution of the conifold singularity\label{bhres}}

The presence of singularities on the \Ka (resp. complex
structure) moduli space of $(2,2)$ superconformal Calabi-Yau sigma-models
where 2-cycles (resp. 3-cycles) vanish has been a long standing
problem in the context of type II superstring compactifications on a CY
threefold
$\C$. The issue is particularly sharp in the case of the
type IIB string at a complex structure singularity,
since the metric on the vector multiplets is exact at tree-level
in perturbation theory, and receives no worldsheet instanton corrections.
The logarithmic singularity in the ${\cal N}=2$ prepotential
signals that some light degrees of freedom have been integrated
out. Indeed, the type IIB theory possesses D3-branes, which
by wrapping the vanishing 3-cycle yield light black hole states
in four dimensions. Those are BPS hypermultiplets charged under 
the $U(1)$ vector associated to the cycle and whose mass is proportional 
to the volume of the vanishing cycle, 
$m \propto V_{\gamma_3}/g_s$, where $g_s$ is the
string coupling constant. These states are massless at the
conifold even at arbitrarily weak coupling and, when
taken into account, yield a smooth low energy effective 
action \cite{Strominger:1995cz}.
The case of singularities in the \Ka structure of type IIA
compactifications is identical, with D2-branes wrapped on
the vanishing 2-cycle playing the role of the massless black holes,
and we shall refer to these two cases as the ``vector-conifold''.
The reversed process, namely singularities in the \Ka structure of 
type IIB compactifications, 
or complex structure of type IIA, occur in the hypermultiplet
moduli space, and is of a different nature from physical point of 
view. We shall refer to
them as the ``hyper-conifold''. In that case, the Euclidean
D-string wrapped on the vanishing 2-cycle (or D2-brane wrapped
on the vanishing 3-cycle) yield space-time instantons, that have been
argued to correct the singular metric on the hypermultiplet moduli 
space \cite{Ooguri:1996me}. At the same time, the wrapped D3 (or D4) yield 
tensionless strings, which provide the missing degrees of 
freedom \cite{Greene:1996dh}.
Note that from the mathematical point of view,
the vector- and hyper-conifold are the two sides of
the conifold transition, since a two-cycle is shrunk to zero size
and reappears as a three-cycle. 

The mathematical transition between Calabi-Yau manifolds can be
understood as Higgsing/un-Higgsing of the low-energy degrees
of freedom as follows \cite{Greene:1995hu} (we phrase our discussion
in terms of the type IIA vector-conifold). Consider a singularity 
in the \Ka moduli space, where $N$ 2-cycles 
$\gamma_a$ $(a=1,...,N)$ go to zero size
simultaneously ($N=8$ in our first example
Eq. (\ref{split1})). In general,  
these distinct cycles are not independent in homology, but 
they satisfy $R$ relations of the form
\begin{equation}
\alpha_1^r \gamma_1 + \dots + \alpha_N^r \gamma_N = 0\quad
(r=1,...,R)\; ,
\label{contrainte}
\end{equation} 
for some integer $\alpha_a^r$ $(a=1,...,N;
r=1,...,R)$. The membranes wrapped around the 2-cycles give $N$ black holes
hypermultiplets, which are charged under the 
$(N-R)$ independent $U(1)$ massless vector fields arising from
the reduction of the 10-dimensional three-form $C$ on the $(N-R)$
homology classes. In all the transitions
considered in the previous section, 
the $N$ vanishing cycles are all proportional in homology to the same
$S^2$, whose volume is sent to zero, so that $R=N-1$.

Due to their charges, the hypermultiplets are no longer decoupled
from the vectors, but couple in a way consistent with $\N=2$ SUSY.
In $\N=1$ terminology, there is a superpotential 
\begin{equation}
{\cal W}=\sum_{I=1}^{N-R} \sum_{a=1}^{N} q_I^a T^I h_a \tilde h_a
\label{superpot}
\end{equation}
together with D-terms for each of the generators of the gauge group,
\begin{equation}
D_I = \sum_{a=1}^{N} q_I^a \left( |h_a|^2 - |\tilde h_a|^2 \right),
\label{Dterm}
\end{equation}
where $T^I$ are the complex scalar fields in the vector multiplets,
and $(h_a,\tilde h_a)$ the chiral fields associated to
the $N$ black holes hypermultiplets, with charge $q_I^a$ under the
$I$-th $U(1)$ factor.
In the Coulomb phase where $T^I$ condense, all hypermultiplets
get a mass, yielding the massless spectrum on the CY $\C$.
In the Higgs phase on the other hand, $3(N-R)$  among the $4N$ real 
scalars degrees of freedom are fixed by the D-term and $T^I$'s F-term
conditions, and another $(N-R)$ is gauged away by
the  $U(1)^{N-R}$ vector fields getting massive. This leaves 
$4R$ real flat directions in the potential corresponding to 
$R$ neutral hypermultiplets. The spectrum in the Higgs phase
consists of $h_{12}+R$ neutral
hypermultiplets together with $h_{11}-(N-R)$ abelian vector multiplets
coupled to ${\cal N}=2$ supergravity. This is the spectrum 
of a compactification on a new CY $\C'$, whose Hodge numbers are
$h'_{12} :=h_{12}+R$ and $h'_{11} :=h_{11}-(N-R)$. This is precisely
what we found by determinantal contraction in all the examples
considered in the previous section, as can be checked by using
Eq. \eqref{euler}.

\section{Conifold transitions between G$_2$ manifolds\label{G2case}}

Having recalled the necessary background on the conifold transition
between Calabi-Yau manifolds, we now discuss the transition between
$G_2$ manifolds constructed by a {\it freely acting}
quotient $(\C\times S^1)/\sigma$. The non-freely acting case
will be discussed in Section 5.
We start by considering the same transition as in Eq. \eqref{split1},
but now keeping track of the antiholomorphic involution $\sigma$.

\subsection{An example\label{G2ex}}
Let us consider a $G_2$ manifold constructed 
from our first example \eqref{M1}. Since the 
embeding space in this case is $\CP^7 \times \CP^1$, we may choose
an involution of type A or B on each factor, thus
giving 4 different orbifolds.
In order to exclude the possibility of fixed points, 
we choose the involution B acting on $\CP^7$.
As we shall see, this action can be combined
with any of the involutions A or B on $\CP^1$ (and also involution C
acting on two 
$\CP^1$'s), to give a  consistent involution $\sigma$.

\subsubsection{Involution $B\times B$ on $\CP^7 \times \CP^1$}
We start by considering the configuration matrix
\begin{equation}
\N^{(1)} =  \left[  \begin{array}{c||ccccc} \sh &1&1&2&2&2 \\ \uh &1&1&0&0&0   \end{array}
\right]^{0,61} \; ,
\end{equation}
where we anticipated the values of $b_2$ and $b_3$ that
will be determined momentarily. The invariance under the
involution $w$: $x_i\rightarrow
\xhb_i$ $(i=1,...,8)$, $y_j\rightarrow \yhb_j$ $(j=1,2)$
puts constraints on the 
polynomials appearing in the system $(\Sy_1)$ of \eqref{eq}.
Indeed, the transformed system under the involution should 
be equivalent to the complex conjugate of the
original one:
\begin{equation}
\left\{ \begin{array}{lll}
f_1(\xhb,\yhb) =  f_2(\xhb,\yhb) = 0\\
e_{2}(\xhb)= e_{3}(\xhb) = e_{4}(\xhb) =0
\end{array} \right.
\quad \Longleftrightarrow\quad
\left\{ \begin{array}{lll}
\fb_1(\xb,\yb) = \fb_2(\xb,\yb) =0 \\
\eb_{2}(\xb)= \eb_{3}(\xb) = \eb_{4}(\xb) =0 \; ,
\end{array} \right.
\label{syseq}
\end{equation}
where $\fb_i(u,v)$ $(i=1,2)$ is the polynomial $f_i(u,v)$, whose
coefficients have been changed to their complex conjugates, while
$\xh_{2p-1}=-x_{2p}$, $\xh_{2p}=x_{2p-1}$ $(p=1,2,3,4)$, and
similarly for $\eb_n$ $(n=2,3,4)$ and $\yh_j$  $(j=1,2)$.  As a
result, there should exist two 
matrices $M$ and $N$ in $GL(2,\Comp)$ and $GL(3,\Comp)$, respectively,
such that
\be 
\pamatrix{ f_1(x,y)  \cr  f_2(x,y)  } = M
\pamatrix{ \fb_1(\xh,\yh)  \cr \fb_2(\xh,\yh) } \quad \mbox{and} \quad
\pamatrix{ e_2(x)  \cr e_3(x) \cr e_4(x) } = N
\pamatrix{ \eb_2(\xh)  \cr \eb_3(\xh)\cr \eb_4(\xh)  }  ,
\end{equation}
where we have used the fact that the dummy variables satisfy $\xhh=-x$,
$\yhh=-y$. 
The consistency of this system imposes 
\be 
\label{MN}
MM^* = I \quad \mbox{and} \quad NN^* = I \; .
\end{equation}
By considering linear changes of basis on the equations $f_{1,2}$ and
$e_{2,3,4}$ similar to whose considered in
Section \ref{invproj} for projective coordinates, it is possible to
choose $M=I$, $N=I$, without loss of generality. As a
result, the coefficients of the equations must satisfy
\begin{equation}
f_i(x,y)=\fb_i(\xh,\yh) \quad (i=1,2)\quad  \mbox{and}\quad
e_j(x)=\eb_j(\xh) \quad 
(j=2,3,4) \; 
\end{equation}
or equivalently in terms of the polynomials $P_{kl}$:
\begin{equation}
\label{condpol}
P_{k1}(x)=\bar{P}_{k2}(\xh) \qquad (k=1,2) \; .
\end{equation}
This condition halves the number of parameters
appearing in the coefficients of the defining polynomials.
This is in agreement with the effect of the orbifold on the untwisted
spectrum discussed in Section \ref{cons}, at least when the 
parameters appearing in the defining equations are 
in one-to-one correspondence with the complex structure
moduli of the CY (see earlier footnote in Section \ref{example}).

The Betti numbers of a $G_2$ manifold $\G_1$ belonging to $\N^{(1)}$
can be determined simply as follows. 
The harmonic $(1,1)$-forms of the CY are the pull-back of the \Ka
forms $J_7$ and $J_1$ of the projective spaces $\CP^7$ and $\CP^1$,
which are odd under the involution.
Hence $h_{11}^+=0$, so that using $h_{11}=2$, $h_{12}=58$ for the CY, we
find from \eqref{sp} that $b_2=0$ and $b_3=1+2+58=61$. Note that this
is the complete spectrum, since there are no fixed points that could
contribute twisted sectors.

We can now perform the determinantal contraction described
in Section \ref{example} by sending the volume $v_1$ of $\CP^1$ to
zero and going to the description in terms of the system (\ref{eqsing}). 
We then note that the relations
\begin{equation}
\label{ehat}
e_1^\sharp (x)=\eb_1^\sharp (\xh)\quad \mbox{and} \quad
e_j(x)=\eb_j(\xh) \quad  
(j=2,3,4) \; 
\end{equation}
hold, thanks to Eq. (\ref{condpol}) and to the fact that $P_{kl}$ are
polynomials of odd degree. This defines a singular variety
$\G_{0}^\sharp$ corresponding to a point in the 
$G_2$ moduli space, with
configuration matrix 
\begin{equation}
\N^{(0)}= \left[  \begin{array}{c||cccc} \sh &2&2&2&2  \end{array}
\right]^{0,67}  .
\end{equation}
Indeed, proceeding as for $\N^{(1)}$, it is easy to show that the
4 generic homogeneous equations of degree 2 in \eqref{eqreg}
have to satisfy the same constraints as in \eqref{ehat}.
$\G_0^\sharp$ may now be deformed into a smooth manifold
on $\N^{(0)}$ by considering some real $t$ and $\varepsilon^2(x)$ in
\eqref{eqreg} such that $\varepsilon^2(x)\equiv \bar{\varepsilon}^2(\xh)$. 
The Betti numbers on $\N^{(0)}$ are found in the same way as above: 
The \Ka form $J_7$ is odd, so that $h_{11}^+=0$, which along with
$h_{11}=1$, $h_{12}=65$ implies
$b_2=0$ and $b_3=67$.
As a result, we have described the conifold transition
obtained by determinantal contraction from $\N^{(1)}$ to
$\N^{(0)}$. This will be denoted:
\begin{equation}
\left[  \begin{array}{c||cccc} \sh &2&2&2&2  \end{array}
\right]^{0,67} 
\leftarrow
\left[  \begin{array}{c||ccccc} \sh &1&1&2&2&2 \\ \uh &1&1&0&0&0   \end{array}
\right]^{0,61}  .
\label{split}
\end{equation}
On the double cover of the orbifold $\G_1$, at each point of $S^1$,
the CY admitting the isometry $B\times B$ have 8 $S^2$'s that have shrunk
to points on $\G_0^\sharp$ before being deformed to $S^3$'s on $\G_0$.

\subsubsection{Involution $B\times     A$ on $\CP^7 \times \CP^1$}
We now consider the configuration matrix
\begin{equation}
\left[  \begin{array}{c||ccccc} \sh &1&1&2&2&2 \\ \ub &1&1&0&0&0   \end{array}
\right]^{0,61} .
\end{equation}
The system $(\Sy_1)$ of Eq. (\ref{eq}) must satisfy
\begin{equation}
\left\{ \begin{array}{lll}
f_1(\xhb,\yb) =f_2(\xhb,\yb) = 0 \\
e_{2}(\xhb)= e_{3}(\xhb) = e_{4}(\xhb) =0
\end{array} \right.
\quad \Longleftrightarrow\quad
\left\{ \begin{array}{lll}
\fb_1(\xb,\yb) = \fb_2(\xb,\yb) =0 \\
\eb_{2}(\xb)= \eb_{3}(\xb) = \eb_{4}(\xb) =0 \; ,
\end{array} \right.
\label{syseq2}
\end{equation}
which implies now that there exist two
matrices $M$ and $N$ in $GL(2,\Co)$ and $GL(3,\Co)$, respectively,
such that
\be 
\pamatrix{f_1(x,y)  \cr f_2(x,y) } = M
\pamatrix{ \fb_1(\xh,y)  \cr \fb_2(\xh,y) } \quad \mbox{and} \quad
\pamatrix{ e_2(x)  \cr e_3(x) \cr e_4(x)  } = N
\pamatrix{ \eb_2(\xh)  \cr \eb_3(\xh)\cr \eb_4(\xh) } .
\end{equation}
Since $f_i(x,y)$ $(i=1,2)$ is of odd degree in $x$, 
Eq. \eqref{MN} is now replaced by 
\be 
\label{MN2}
MM^* = -I \quad \mbox{and} \quad NN^* = I \; .
\end{equation}
By considering changes of basis in the defining equations, we may
impose without 
loss of generality as in the previous section $M=\pamatrix{ 0&1\cr -1&0 }$ and $N=I$:
\begin{equation}
f_1(x,y)=\fb_2(\xh,y) \quad \mbox{and} \quad e_j(x)=\eb_j(x)\quad
  (j=2,3,4)
\end{equation}
or equivalently
\begin{equation}
\label{condpol2}
P_{1l}(x)=\bar{P}_{2l}(\xh) \qquad (l=1,2) \; .
\end{equation}

The Betti numbers of the $G_2$ orbifold can be computed in the
same way as before, yielding $b_2=0$, $b_3=61$. 
Upon sending the volume of $\CP^1$ to zero, the system
becomes $(\Sy_0^\sharp)$ in Eq. (\ref{eqsing}), which satisfies
Eq. \eqref{ehat}, as can be seen from Eq. (\ref{condpol2}). 
This shows that we arrived at a singular point of the
moduli space $\N^{(0)}$. As before, we can then deform the orbifold to
obtain a smooth manifold of $\N^{(0)}$. Thus, we find another
transition, whose end point is in $\N^{(0)}$:
\begin{equation}
\left[  \begin{array}{c||cccc} \sh &2&2&2&2  \end{array}
\right]^{0,67} 
\leftarrow
\left[  \begin{array}{c||ccccc} \sh &1&1&2&2&2 \\ \ub &1&1&0&0&0   \end{array}
\right]^{0,61}  .
\label{split'}
\end{equation}

\subsubsection{Involution $B\times C$ on $\CP^7 \times \CP^1\times \CP^1$}
So far, all the $G_2$ manifolds that we constructed had
$b_2=0$. This is because 
the K{\"a}hler forms $J_{7,1}$  were odd so
that $h_{11}^+$ was always zero. We are going to see now that this is
not a general feature, when one uses involutions $C$. As an example, let
us consider the configuration matrix 
\begin{equation}
\left[  \begin{array}{c||cccccc} \sh &1&1&1&1&2&2 \\ \uf &1&1&0&0&0&0
    \\ \us &0&0&1&1&0&0 \end{array} 
\right]^{1,54} ,
\label{cc}
\end{equation}
where $b_2$ will turn out to be one. The corresponding system of
equations
\begin{equation}
(\Sy_2)\left\{ \begin{array}{l}
f_1(x,y) := P_{11}(x)y_1+P_{12}(x)y_2 =0 \\
f_2(x,y) := P_{21}(x)y_1+P_{22}(x)y_2 =0 \\
f_3(x,z) := P_{31}(x)z_1+P_{32}(x)z_2 =0 \\
f_4(x,z) := P_{41}(x)z_1+P_{42}(x)z_2 =0 \\
e_{3}(x)= e_{4}(x) =0 \; ,
\end{array} \right.
\label{eq2}
\end{equation}
where $P_{k,l}$ $(k=1,...,4\; ; \; l=1,2)$ and $e_n$ $(n=3,4)$ are
polynomials in $x_i$'s of degree 1 and 2, respectively, while $z_j$
$(j=1,2)$ are projective coordinates for the second $\CP^1$
factor. For this system to define a manifold satisfying the 
discret isometry $B$ on $\CP^7$, $C$ on $\CP^1 \times \CP^1$, we must have
\begin{equation}
\left\{ \begin{array}{lll}
f_1(\xhb,\yb) = f_2(\xhb,\yb) = 0 \\
f_3(\xhb,\zb) = f_4(\xhb,\zb) = 0 \\
e_{3}(\xhb) = e_{4}(\xhb) =0 \; ,
\end{array} \right.
\quad \Longleftrightarrow\quad
\left\{ \begin{array}{lll}
\fb_1(\xb,\zb) =\fb_2(\xb,\zb) =0 \\
\fb_3(\xb,\yb) =\fb_4(\xb,\yb) =0 \\
\eb_{3}(\xb) = \eb_{4}(\xb) =0 \; ,
\end{array} \right.
\label{syseq3}
\end{equation}
which shows that their exist $M,M'$ and $N$ in $GL(2,\Co)$ such that
\begin{eqnarray}
&\pamatrix{f_1(x,y)  \cr f_2(x,y) } = M
\pamatrix{ \fb_3(\xh,y)  \cr \fb_4(\xh,y) } \; ,\quad
\pamatrix{f_3(x,z)  \cr f_4(x,z) } = M'
\pamatrix{ \fb_1(\xh,z)  \cr \fb_2(\xh,z) }\nonumber \\\label{3eq} \\ 
&\mbox{and} \quad
\pamatrix{ e_3(x) \cr e_4(x)  } = N
\pamatrix{ \eb_3(\xh)\cr \eb_4(\xh) }.\nonumber
\end{eqnarray}
The consistency of this system implies that
$M{M'}^*=-I$ and $NN^*=I$, so that we may choose
without loss of generality $M=-M'=I, N=I$.
As a
result, the configuration matrix in Eq. (\ref{cc}) corresponds to the system
$(\Sy_2)$ in Eq. (\ref{eq2}), with the constraints
\begin{equation}
f_k(x,y)=\fb_{k+2}(\xh,y) \quad (k=1,2)\quad \mbox{and} \quad
e_j(x)=\eb_j(x)\quad (j=3,4)\; ,
\end{equation}
which gives 
\begin{equation}
P_{kl}(x)=\bar{P}_{k+2,l}(\xh) \quad (k,l=1,2)\; . 
\label{pkl}
\end{equation}

To determine the Betti numbers, denoting by 
$J_7,J_1,J_1'$ the \Ka forms of the three projective spaces,
we note that $J_7$ and $J_1+J_1'$ are still odd, but 
$J_1-J_1'$ is even. The volume of the two $\CP^1$ are
therefore restricted to be the same, but their common volume
is free to vary. Hence $h_{11}^+=1$,
and since $h_{11}= 3$, $h_{12}=51$, we find
$b_2=1$, $b_3=1+2+51=54$. These are the exact values of the
Betti numbers, since again there are no fixed points that could contribute
twisted sectors.

Let us now replace the equations
$f_i(x,y)=0$ $(i=2,4)$ in Eq. (\ref{eq2}) 
by the vanishing determinants
\begin{equation}
\begin{array}{l}
e_1^\sharp (x) = P_{11}(x)P_{22}(x)-P_{21}(x)P_{12}(x)=0 \\ 
e_2^\sharp (x)= P_{31}(x)P_{42}(x)-P_{41}(x)P_{32}(x)=0\;  
\end{array}
\end{equation}
and send the volume of both $\CP^1$ to zero, so that the system
becomes   
\begin{equation}
\left\{
\begin{array}{l}
e_1^\sharp (x)=e_2^\sharp(x)=0\\
e_3(x)=e_4(x)=0\;.
\end{array} 
\right.
\label{eqsing2}
\end{equation}
We then note that thanks to Eq. (\ref{pkl}), we have 
\begin{equation}
\pamatrix{ e_1^\sharp(x) \cr e_2^\sharp(x)  } = N
\pamatrix{ \eb_1^\sharp(\xh)\cr \eb_2^\sharp(\xh) }\; ,\quad \mbox{where}
\quad N=\pamatrix{ 0 & 1 \cr 1 & 0}\; .
\label{ehat2}
\end{equation}
Since $NN^*=I$, we could by a change of basis in the 2
dimensional vectorial space of equations $e_{1,2}^\sharp$
replace $N$ by the identity matrix, as we did for $e_{3,4}$ in
Eq. (\ref{3eq}). As a 
result, the equations in (\ref{eqsing2}) define a singular space in
$\N^{(0)}$ that we can deform to a smooth manifold by adding $t_i
\varepsilon_i^2(x)$ $(i=1,2)$ to the right hand sides of the two first
equations, where $t_{1,2}$ are real numbers and $\varepsilon^2_{1,2}$
are generic homogeneous polynomials of degree two satisfying
$\varepsilon_1^2(x)=\bar{\varepsilon}_2^2(\xh)$. This is summarized as
\begin{equation}
\left[  \begin{array}{c||cccc} \sh &2&2&2&2  \end{array}
\right]^{0,67} 
\leftarrow
\left[  \begin{array}{c||cccccc} \sh &1&1&1&1&2&2 \\ \uf &1&1&0&0&0&0
    \\ \us &0&0&1&1&0&0 \end{array}
\right]^{1,54} .
\end{equation}

\subsection{The web of $G_2$ manifolds $(\mbox{\rm CICY} \times S^1)/\sigma$}
\label{G2web}
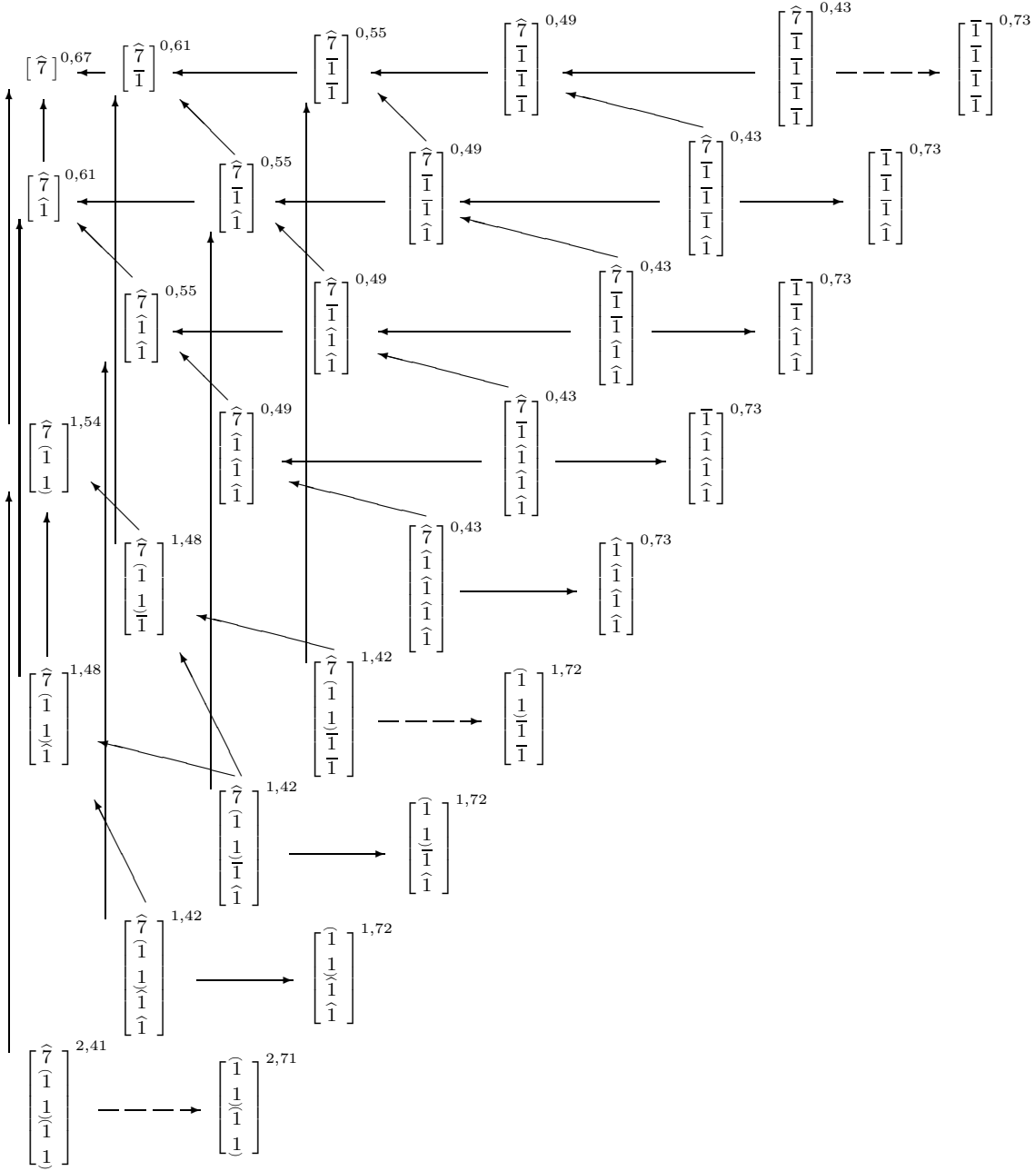
\begin{figure}
\setlength{\unitlength}{1cm}
\begin{picture}(15,18)
\put(0,9){
{\scriptsize $
\begin{array}{lllllllllll}
\left[ \begin{array}{l} \sh \end{array} \right]^{0,67} & \left[
    \begin{array}{l} 
    \sh \\ \ub \end{array} \right]^{0,61} &   &  \left[ \begin{array}{l}
    \sh \\ \ub \\ \ub \end{array} \right]^{0,55} &  & \left[ \begin{array}{l}
    \sh \\ \ub \\ \ub \\ \ub \end{array} \right]^{0,49} &  &  & \left[
    \begin{array}{l} 
    \sh \\ \ub \\ \ub \\ \ub \\ \ub \end{array} \right]^{0,43} &   &\left[
    \begin{array}{l} 
    \ub \\ \ub \\ \ub \\ \ub \end{array} \right]^{0,73}\\
\left[ \begin{array}{l} \sh \\ \uh \end{array} \right]^{0,61} &  &  \left[
    \begin{array}{l} \sh \\ \ub \\ \uh \end{array} \right]^{0,55} &  & \left[
    \begin{array}{l} \sh \\ \ub \\ \ub \\ \uh \end{array} \right]^{0,49} &  & 
    & \left[ \begin{array}{l} \sh \\ \ub \\ \ub \\ \ub \\ \uh
    \end{array} \right]^{0,43}&  & \left[ \begin{array}{l}  \ub \\ \ub
      \\ \ub \\ \uh 
    \end{array} \right]^{0,73} \\ 
  & \left[ \begin{array}{l} \sh \\ \uh \\ \uh \end{array} \right]^{0,55} &  &
    \left[ \begin{array}{l} \sh \\ \ub \\ \uh \\ \uh \end{array}
    \right]^{0,49} &  &  & \left[ \begin{array}{l} \sh \\ \ub \\ \ub \\ \uh \\
    \uh \end{array} \right]^{0,43}&  & \left[ \begin{array}{l}  \ub \\ \ub
    \\ \uh \\ 
    \uh \end{array} \right]^{0,73} \\ 
\left[ \begin{array}{l} \sh \\ \ufp \\ \usp \end{array} \right]^{\!\!1,54} &  &
    \left[ \begin{array}{l} \sh \\ \uh \\ \uh \\ \uh \end{array}
    \right]^{0,49} &  &  & \left[ \begin{array}{l} \sh \\ \ub \\ \uh \\ \uh \\
    \uh \end{array} \right]^{0,43}&  & \left[ \begin{array}{l} \ub \\ \uh
    \\ \uh \\ 
    \uh \end{array} \right]^{0,73} \\ 
  & \left[ \begin{array}{l} \sh \\ \ufp \\ \usp \\ \ub \end{array}
    \right]^{1,48} &  &  & \left[ \begin{array}{l} \sh \\ \uh \\ \uh \\ \uh \\
    \uh \end{array} \right]^{0,43}&  & \left[ \begin{array}{l} \uh \\ \uh
    \\ \uh \\ 
    \uh \end{array} \right]^{0,73} \\ 
\left[ \begin{array}{l} \sh \\ \ufp \\ \usp \\ \uh \end{array}
    \right]^{\!\!1,48} &  &  & \left[ \begin{array}{l} \sh \\ \ufp \\ \usp \\ \ub \\
    \ub \end{array} \right]^{1,42}&  & \left[ \begin{array}{l}  \ufp \\ \usp
    \\ \ub \\ 
    \ub \end{array} \right]^{1,72} \\ 
  &  & \left[ \begin{array}{l} \sh \\ \ufp \\ \usp \\ \ub \\
    \uh \end{array} \right]^{1,42}&   &\left[ \begin{array}{l}  \ufp \\ \usp
    \\ \ub \\ 
    \uh \end{array} \right]^{1,72} \\
  & \left[ \begin{array}{l} \sh \\ \ufp \\ \usp \\ \uh \\
    \uh \end{array} \right]^{1,42}&   &\left[ \begin{array}{l}  \ufp \\ \usp
    \\ \uh \\ 
    \uh \end{array} \right]^{1,72} \\
\left[ \begin{array}{l} \sh \\ \ufp \\ \usp \\ \ufp \\
    \usp \end{array} \right]^{2,41} &  & \left[ \begin{array}{l}  \ufp \\
    \usp \\ \ufp \\ 
    \usp \end{array} \right]^{2,71}
\end{array}
$}
}
\put(1.4,16.6){\vector(-1,0){.4}}
\put(4.2,16.6){\vector(-1,0){1.8}}
\put(7,16.6){\vector(-1,0){1.7}}
\put(10.9,16.6){\vector(-1,0){2.8}}
\put(12.1,16.6){\line(1,0){.3}}\put(12.5,16.6){\line(1,0){.3}}
\put(12.9,16.6){\line(1,0){.3}} \put(13.3,16.6){\vector(1,0){.3}} 

\put(2.7,14.7){\vector(-1,0){1.7}}
\put(5.5,14.7){\vector(-1,0){1.6}}
\put(9.5,14.7){\vector(-1,0){2.9}}
\put(10.7,14.7){\vector(1,0){1.5}}

\put(4,12.8){\vector(-1,0){1.6}}
\put(8.2,12.8){\vector(-1,0){2.8}}
\put(9.4,12.8){\vector(1,0){1.5}}

\put(6.9,10.9){\vector(-1,0){2.9}}
\put(8.0,10.9){\vector(1,0){1.6}}

\put(0.5,15.3){\vector(0,1){.9}}
\put(3.3,15.4){\vector(-1,1){.8}}
\put(6.1,15.6){\vector(-1,1){.7}}
\put(10.15,15.8){\vector(-4,1){2}}

\put(1.8,13.6){\vector(-1,1){.8}}
\put(4.6,13.7){\vector(-1,1){.7}}
\put(8.9,13.9){\vector(-4,1){2.3}}

\put(3.2,11.8){\vector(-1,1){.7}}
\put(7.3,12.0){\vector(-4,1){1.9}}

\put(1.9,9.9){\vector(-1,1){.7}}
\put(6.1,10.1){\vector(-4,1){2.0}}
\put(6.6,9.0){\vector(1,0){1.7}}

\put(.55,8.05){\vector(0,1){2.1}}
\put(4.75,8.15){\vector(-4,1){2.0}}
\put(5.4,7.1){\line(1,0){.3}}\put(5.8,7.1){\line(1,0){.3}}
\put(6.2,7.1){\line(1,0){.3}} \put(6.6,7.1){\vector(1,0){.3}} 

\put(3.3,6.3){\vector(-4,1){2.0}}
\put(3.4,6.3){\vector(-1,2){.9}}
\put(4.1,5.15){\vector(1,0){1.4}}

\put(2.0,4.45){\vector(-1,2){.75}}
\put(2.75,3.3){\vector(1,0){1.4}}

\put(1.3,1.4){\line(1,0){.3}}\put(1.7,1.4){\line(1,0){.3}}
\put(2.1,1.4){\line(1,0){.3}} \put(2.5,1.4){\vector(1,0){.3}} 

\put(0.0,11.45){\vector(0,1){4.9}}
\put(0.0,2.25){\vector(0,1){8.2}}
\put(0.15,7.75){\vector(0,1){6.7}}

\put(1.40,4.2){\vector(0,1){8.15}}
\put(1.55,9.7){\vector(0,1){6.55}}

\put(2.95,6.1){\vector(0,1){8.15}}

\put(4.35,7.95){\vector(0,1){8.20}}

\end{picture}
\caption{Sequences of transitions relating the 9 $G_2$ configuration matrices
descending from  $\M^{(1111)}$. Only the first 
column of the $G_2$ configuration matrices is shown, the other ones
being identical to those appearing in the CY configuration matrices of
Eq. (\ref{seq}). Dotted arrows denote transitions
to non-freely acting orbifolds.}
\label{fig}
\end{figure}

\begin{figure}
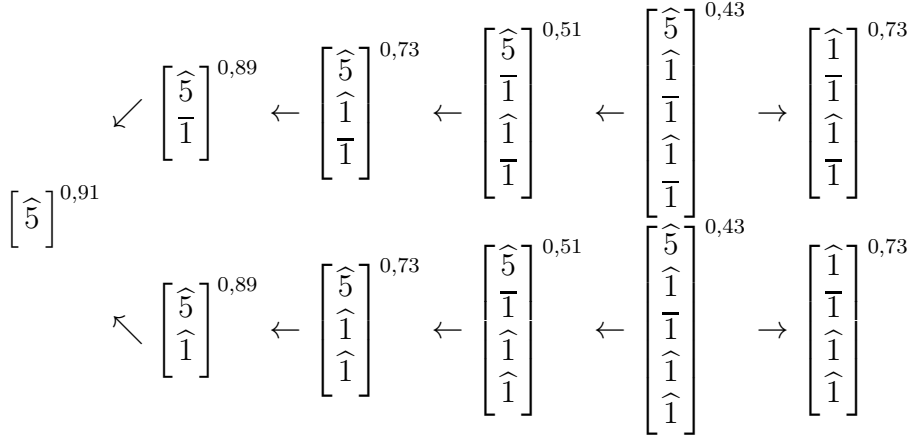

\begin{equation*}
\left[  \begin{array}{c} \ch   \end{array}
\right]^{0,91} 
\begin{array}{llllllllll}
\swarrow&
\left[  \begin{array}{c} \ch  \\ \ub    \end{array}
\right]^{0,89}
&\leftarrow&
\left[  \begin{array}{c} \ch  \\ \uh \\  \ub  \end{array}
\right]^{0,73}
&\leftarrow&
\left[  \begin{array}{c} \ch  \\ \ub 
    \\  \uh \\  \ub \end{array}
\right]^{0,51}
&\leftarrow&
\left[  \begin{array}{c} \ch  \\ \uh 
    \\  \ub  \\  \uh \\ \ub \end{array}
\right]^{0,43}
&\rightarrow&
\left[  \begin{array}{c}
                \uh \\
                \ub \\
                \uh \\
                \ub 
                \end{array}     \right]^{0,73}\\
\nwarrow&
\left[  \begin{array}{c} \ch  \\ \uh    \end{array}
\right]^{0,89}
&\leftarrow&
\left[  \begin{array}{c} \ch  \\ \uh \\  \uh  \end{array}
\right]^{0,73}
&\leftarrow&
\left[  \begin{array}{c} \ch  \\ \ub 
    \\  \uh \\  \uh \end{array}
\right]^{0,51}
&\leftarrow&
\left[  \begin{array}{c} \ch  \\ \uh 
    \\  \ub  \\  \uh \\ \uh \end{array}
\right]^{0,43}
&\rightarrow&
\left[  \begin{array}{c}
                \uh \\
                \ub \\
                \uh \\
                \uh 
                \end{array}     \right]^{0,73}.
\end{array} 
\label{seq'g2}
\end{equation*}
\caption{Sequences of transitions descending from the CICY
sequence \eqref{seq'}. Only the first column of the matrices is shown, the
other are as in (\ref{seq'}).}
\label{fig2}
\end{figure}

As we have seen in Section \ref{CYweb}, all complete intersection CY moduli
spaces are connected to $\M^{(1111)}$, the moduli space associated to
the last configuration matrix in Eq. (\ref{seq}). As a result, $G_2$ manifolds
constructed by orbifolding a product $\C\times S^1$, where $\C$ is
a CICY may also be connected to one of the $G_2$ manifolds
descending from $\M^{(1111)}$. There are 9 possible choices of 
antiholomorphic involutions, 6 of them involve at least one involution
$B$ on a projective factor and are freely acting:
\begin{equation}
\left[  \begin{array}{c||c}
                \uh&2 \\
                \uh&2 \\
                \uh&2 \\
                \uh&2 
                \end{array}     \right]^{0,73} ,
\left[  \begin{array}{c||c}
                \uh&2 \\
                \uh&2 \\
                \uh&2 \\
                \ub&2 
                \end{array}     \right]^{0,73},
\left[  \begin{array}{c||c}
                \uh&2 \\
                \uh&2 \\
                \ub&2 \\
                \ub&2 
                \end{array}     \right]^{0,73},
\left[  \begin{array}{c||c}
                \uh&2 \\
                \ub&2 \\
                \ub&2 \\
                \ub&2 
                \end{array}     \right]^{0,73},
\left[  \begin{array}{c||c}
                \uh&2 \\
                \uh&2 \\
                \uf&2 \\
                \us&2 
                \end{array}     \right]^{1,72},
\left[  \begin{array}{c||c}
                \uh&2 \\
                \ub&2 \\
                \uf&2 \\
                \us&2 
                \end{array}     \right]^{1,72},
\label{1111nf}
\end{equation}
while the last 3 have fixed points,
\begin{equation}
\left[  \begin{array}{c||c}
                \ub&2 \\
                \ub&2 \\
                \ub&2 \\
                \ub&2 
                \end{array}     \right]^{0,73} ,
\left[  \begin{array}{c||c}
                \ub&2 \\
                \ub&2 \\
                \uf&2 \\
                \us&2 
                \end{array}     \right]^{1,72},
\left[  \begin{array}{c||c}
                \uf&2 \\
                \us&2 \\
                \uf&2 \\
                \us&2 
                \end{array}     \right]^{2,71}.
\label{1111f}
\end{equation}
Here the indicated Betti numbers do not take into account the contributions 
from the twisted sectors. This is not to say that the web of $G_2$
manifolds descending 
from CICY splits into 9 disconnected components. Indeed, it is easy to
find sequences of determinantal splittings and contractions 
that relate any of these 9 cases, as shown in Figure \ref{fig}.
In the sequences we displayed, the absence of fixed points
is ensured by taking an involution of type $B$ on the $\CP^7$ factor.
This rule is broken only when arriving to the three 
non-freely acting configurations in Eq. \eqref{1111f}, shown with dotted
arrows on Figure \ref{fig}. It would be interesting
to understand better these transitions with fixed points.
For illustrative purposes, we also give in Figure 
\ref{fig2} the sequences of $G_2$
manifolds descendings the CY sequence \eqref{seq'} using the
involution $B$ on $\CP^5$, so that the involution is
freely acting. 
It is easy to check that the sequences shown are the only
ones for which the involution is compatible with the 
CICY matrix, as discussed in Section \ref{nota}.
Although the above sequences are suggestive, we have not shown 
that all  $G_2$ manifolds (CICY$\times
S^1)/\sigma$ could be related to one of the 9 configuration matrices above, 
since intermediate 
steps could in principle involve non-freely acting configurations.

Finally, we note that several of the manifolds in Figure 1 
have the same value of $b_2+b_3$. As we mentioned in the introduction, 
this is a necessary condition for $G_2$ manifolds to be mirror to
each other, and it would be interesting to check if this is indeed
the case.

\subsection{Black hole condensation in $\N=1$ vacua}
Having described conifold transitions between  complete intersection
$G_2$ manifolds 
at the mathematical level, we now would like to understand 
these processes in string or field theory terms. As long as the nodal
points are not fixed under the involution, the physical mechanisms will
be very similar to the standard conifold case for CY manifolds in $\N
=2$ type II constructions.
In \cite{Sen:1996tz}, $\mathbb{Z}_2$ orbifolds of
M-theory on $K_3\times S^1$ were considered, breaking $\N=2$ to $\N=1$
in 6 dimensions: In some cases, even though the additional states
arising at specific 
points in moduli space were not BPS any more, they still inherited
their mass formula from the $\N=2$ theory, so that they were still
massless after orbifolding. Similar arguments apply in four dimensions.   
The precise mechanism will actually turn out to be somewhat different
depending whether $b_2$ remains constant or decreases.

\subsubsection{Transition at constant $b_2$}
This situation arises when the $(N-R)$ homology classes of the $N$
2-cycles  vanishing at the singularity 
of the CY are odd under the involution $w$. Accordingly,
the $(N-R)$ $U(1)$ gauge fields present in the $\N=2$ case are projected out,
while the volumes of the cycles together with the B-fluxes 
remain as chiral fields of the $\N=1$ theory. As the
2-cycles shrink to zero size and grow into 3-cycles, no
extra gauge fields appear, so that $b_2$ remains constant.

Since by assumption the antiholomorphic involution has no fixed point,
it maps each node $a$ to a different node $w(a)$ (in particular,
the number of nodes $N$ has to be even).
Since for the cycles $\gamma_a\rightarrow -\gamma_{w(a)}$, the
Ramond-Ramond charged black hole fields have to satisfy the projection
relation
\begin{equation}
h_{a}(x)=\tilde h_{w(a)}(-x)\ ,\quad
\tilde h_{a}(x)=h_{w(a)}(-x)\ ,
\end{equation}
where $x$ is the coordinate along $S^1$. Ordering the $N$ nodal
points so that $w(a)=a+N/2$, the even combinations
$H_a=h_a+\tilde h_{a+N/2}$ and $\tilde H_a=\tilde h_{a}+ h_{a+N/2}$ 
therefore remain massless, while the odd combinations
$h_a-\tilde h_{a+N/2}$ and $\tilde h_{a}- h_{a+N/2}$ 
are projected out, or rather acquire a Kaluza-Klein mass
$1/2R$. The
superpotential \eqref{superpot} therefore  reduces to
\begin{equation}
{\cal W}= \sum_{I=1}^{N-R}\sum_{a=1}^{N/2} q_I^a T^I H_a \tilde H_a\;,
\end{equation}
with usual diagonal quadratic kinetic terms.
In the phase where the complex scalar fields $T^I$ condense, all black hole 
fields $H_a,\tilde H_a$ acquire a mass, and we find the massless
spectrum corresponding 
to the original $G_2$ manifold $\G$. If on the other hand the
black hole fields $H_a, \tilde H_a$ condense, the $(N-R)$ chiral fields $T^I$
acquire a mass. 
The counting of the massless spectrum goes as follows: To
the $b_3$ chiral multiplets at a generic point of the moduli space, we add
$N$ ones associated to the $H_a,\tilde H_a$, substract
$(N-R)$ from F-terms associated to each of the $T^I$'s  
and another $(N-R)$ from the $T^I$ 
that become massive. As a result, we find a new branch of
flat directions of complex dimension $b_3 + 2R-N$. This spectrum
corresponds to a compactification on a $G_2$ manifold, whose Betti
numbers $b'_2$, $b_3'$ satisfy
\begin{equation}
b_2'=b_2\quad \mbox{and} \quad  b_3'=b_3 +2R-N\; .
\label{newBetti}
\end{equation}
In the examples we considered explicitly, since we had $R=N-1$, we
have $b_3'=b_3 + N-2$, reproducing the Betti numbers of the
examples we considered with $b_2$ constant, as can be checked using
Eq. (\ref{euler}). The transition between $G_2$ manifolds at
constant $b_2$ is therefore realized physically by the
transition between two branches of $\N=1$ vacua.

\subsubsection{Transition at decreasing $b_2$}

This situation occurred in the examples we considered
when some $\CP^n\times (\CP^n)'$ pairs of factors acted upon
by involutions C were shrunk to zero size. Let $N$ be
the number of nodal points arising from the vanishing of the 
\Ka moduli of $\CP^n$. Under the involution $w$, 
these are exchanged with the $N$ nodal points arising
from the vanishing of the other projective factor. 
By the same token as above, the $2N$ black hole hypermultiplets
of the $\N=2$ theory reduce to $2N$ chiral multiplets $H_a,\tilde H_a$.
On the other hand, the $2(N-R)$ $\N=2$ vector multiplets reduce
to $(N-R)$ $\N=1$ vector multiplets (from the even homology)
and $(N-R)$ chiral multiplets (from the odd homology). The latter
are neutral under the gauge group $U(1)^{N-R}$, while the
black hole states have charge $\pm q^I_a$. The scalars in this
theory interact through the $\N=1$ superpotential
\begin{equation}
{\cal W} = \sum_{a=1}^{N}  \sum_{I=1}^{N-R} q_a^I T^I H_a \tilde H_a \; ,
\end{equation}
and the D-terms
\begin{equation}
D^I= \sum_{a=1}^{N}  q_a^I \left( |H_a|^2-| \tilde H_a|^2 \right).
\end{equation}
In the Coulomb phase, where the $T^I$ condense and give
a mass to the charged scalars $H_a,\tilde H_a$, we find the expected
massless spectrum of the compactification on the $G_2$ manifold $\G$.
In the Higgs phase, where $H_a,\tilde H_a$ acquire an expectation 
value, the gauge group is Higgsed and the $T^I$ acquire a mass.
The counting of massless chiral fields goes as follows:
We start with $b_3+2N$ chiral fields, impose $(N-R)$ F-term conditions
associated to each of the $T^I$'s as well as
$(N-R)$ real D-term conditions, gauge fix $(N-R)$ real broken generators
and finally give a mass to the $(N-R)$ $T^I$ fields.
The resulting spectrum is therefore that of a compactification
on a new $G_2$ manifold $\G'$ with Betti numbers
\begin{equation}
b_2'= b_2 - (N-R)\quad \mbox{and} \quad  b_3'=b_3 +3R-N\; .
\label{renewBetti}
\end{equation} 
These Betti numbers are precisely what we found from determinantal
contraction in the cases where involutions C were considered, as can
be checked from Eq. (\ref{euler}), renaming $N$ into $2N$ and taking
$R=N-1$. We see that this situation is similar to what 
was already happening in the $\N=2$ case, namely the transition from
one moduli space to another is a realization of a Higgs mechanism.     

\section{Fixed points and conifolds}
Our discussion of conifold transitions in $G_2$ manifolds has so 
far been restricted to the case where the antiholomorphic involution
acts without fixed points. 
A simple way to achieve this was to
choose the freely-acting involution $B$ for one of the
factors of the projective space. 
Clearly, even if the involution has fixed points, the
mechanism of the conifold transitions in $G_2$ manifolds
will remain the same as long as the nodal point themselves
are not fixed under the involution, but exchanged with
one another. On the other hand, when a nodal point is fixed
by the antiholomorphic involution, the local geometry changes
drastically, and new phenomena can be expected. Here we will
content ourselves with a simple example, leaving a more
thorough inverstigation to future work.

Let us consider the local conifold geometry
\begin{equation}
\label{conr}
z_1^2+z_2^2+z_3^2+z_4^2=\epsilon
\end{equation}
and the type $A$ involution $w$: $z_i \to \bar z_i$, 
requiring $\epsilon\in\Real$. Writing $z_i=x^i+i y_i$,
we recognize the defining equations $(x^i)^2-y_i^2=\epsilon,\ x^i y_i=0$
for the cotangent bundle $T^* S^3$. 
For $\epsilon <0$, there is no real solution to 
equation \eqref{conr}, so that the orbifold 
$(T^*S^3 \times \Real )/w\I$, where $\I$: $x\rightarrow -x$ on $\Real$
defines a smooth  
non-compact $G_2$ manifold.
For $\epsilon>0$ on the other hand,
there is a non-empty fixed point set, namely the zero section $S^3$
of the bundle $T^*S^3$. The quotient
$(T^*S^3 \times\Real)/w\I$ now has a conical singularity, locally
$S^3 \times \Real^4/\Zint_2$. Since $b_1(S_3)=0$, the singularity
cannot be resolved so as to preserve $G_2$ holonomy.
The moduli space is therefore restricted to $\epsilon\leq 0$,
with a $SU(2)$ enhanced symmetry point at $\epsilon=0$ coming
from membranes wrapping the $\Real^4/\Zint_2$ singularity.
At $\epsilon=0$, the collapsed 3-cycle may be grown up into a 2-cycle,
changing the topology of the Calabi-Yau threefold to
an $O(-1)\times O(-1)$ bundle over $S^2$, described by
\begin{equation}
\begin{pmatrix} z_1+i z_2 & z_3+i z_4\\-z_3+iz_4 & z_1-iz_2\end{pmatrix}
\begin{pmatrix} \zeta_1 \\ \zeta_2 \end{pmatrix}=0\, .
\end{equation}
The involution $w$ now has to act as $(\zeta_1,\zeta_2)\to
(-\bar\zeta_2,\bar\zeta_1)$ in order to leave this equation invariant,
and therefore is freely acting. We therefore have a smooth $G_2$ manifold
on this side as well. It would be interesting to understand this
phase structure from the point of view of the $SU(2)$ gauge theory
living at the singularity. In particular, the absence of $G_2$
resolution at $\epsilon>0$ may correspond to spontaneous supersymmetry
breaking, and realize a dual version of the scenario proposed in
\cite{Kachru:2000vj} in the context of $\N=1$ theories on type A branes. 
More generally, one may consider 
cases where the special Lagrangian 3-cycle of fixed points
undergoes transitions of the type considered by Joyce \cite{Joyce:1999tz}, 
so that the resulting $G_2$ manifold experiences a topology change.

\acknowledgments
H.P. is indebted to A. Kehagias for his participation at early stages
on related issues. The authors extend warm  thanks to T. H{\"u}bsch
for his explanations on the computation of Hodge numbers in CICY
manifolds. They are also grateful to P. Candelas, P. Kaste, R. Minasian and
C. Schweigert for useful discussions.  
They are both indebted to CERN Theory Division for its kind
hospitality and support, while this work was initiated. 
This work is supported in part by the David and Lucile Packard
foundation and the European  networks FMRX-CT96-0012,
FMRX-CT96-0045, FMRX-CT96-0090, HPRN-CT-2000-00122 and HPRN-CT-2000-00148.


\begin{thebibliography}{cc}

\bibitem{Candelas:1990ug}
P.~Candelas, P.~S.~Green and T.~H{\"u}bsch,
``Rolling Among Calabi-Yau Vacua,''
Nucl.\ Phys.\  {\bf B330}, 49 (1990); ``Connected Calabi-Yau
compactifications (other worlds are just around 
the corner),'' 
{\it Presented at Strings 1988 Meeting, College Park, MD, May 24-28,
  1988}.

\bibitem{Green:1988wa}
P.~S.~Green and T.~H{\"u}bsch,
``Phase transitions among (many of) Calabi-Yau compactifications,''
Phys.\ Rev.\ Lett.\  {\bf 61} (1988) 1163.

\bibitem{Green:1988bp}
P.~S.~Green and T.~H{\"u}bsch,
``Connecting moduli spaces of Calabi-Yau threefolds,''
Commun.\ Math.\ Phys.\  {\bf 119} (1988) 431.

\bibitem{Candelas:1989di}
P.~Candelas, P.~S.~Green and T.~H{\"u}bsch,
``Finite distances between distinct Calabi-Yau vacua: (other worlds
are just around the corner),'' 
Phys.\ Rev.\ Lett.\  {\bf 62} (1989) 1956.

\bibitem{lefschetz}
S. Lefschetz, ``L'analysis situs et la g{\'e}om{\'e}trie alg{\'e}brique,''
Gauthier-Villars, Paris, 1924; reprinted in Selected Papers, Chelsea,
New-York, 1971, pp. 283-439.

\bibitem{Strominger:1995cz}
A.~Strominger,
``Massless black holes and conifolds in string theory,''
Nucl.\ Phys.\  {\bf B451}, 96 (1995)
[hep-th/9504090].

\bibitem{Greene:1995hu}
B.~R.~Greene, D.~R.~Morrison and A.~Strominger,
``Black hole condensation and the unification of string vacua,''
Nucl.\ Phys.\  {\bf B451} (1995) 109
[hep-th/9504145].

\bibitem{Dine:1986zy}
M.~Dine, N.~Seiberg, X.~G.~Wen and E.~Witten,
``Nonperturbative effects on the string world sheet,''
Nucl.\ Phys.\  {\bf B278}, 769 (1986).

\bibitem{Witten:1996bn}
E.~Witten,
``Non-perturbative superpotentials in string theory,''
Nucl.\ Phys.\  {\bf B474}, 343 (1996)
[hep-th/9604030].

\bibitem{Vafa:1996gm}
C.~Vafa and E.~Witten,
``Dual string pairs with N = 1 and N = 2 supersymmetry in four  dimensions,''
Nucl.\ Phys.\ Proc.\ Suppl.\  {\bf 46} (1996) 225
[hep-th/9507050].

\bibitem{Donagi:1996yf}
R.~Donagi, A.~Grassi and E.~Witten,
``A non-perturbative superpotential with E(8) symmetry,''
Mod.\ Phys.\ Lett.\  {\bf A11}, 2199 (1996)
[hep-th/9607091].


\bibitem{Brunner:2000jq}
I.~Brunner, M.~R.~Douglas, A.~Lawrence and C.~Romelsberger,
``D-branes on the quintic,''
JHEP {\bf 0008}, 015 (2000)
[hep-th/9906200].


\bibitem{Kachru:2000ih}
S.~Kachru, S.~Katz, A.~Lawrence and J.~McGreevy,
``Open string instantons and superpotentials,''
Phys.\ Rev.\  {\bf D62} (2000) 026001
[hep-th/9912151].

\bibitem{Kachru:2000an}
S.~Kachru, S.~Katz, A.~Lawrence and J.~McGreevy,
``Mirror symmetry for open strings,''
hep-th/0006047.

\bibitem{Becker:1995kb}
K.~Becker, M.~Becker and A.~Strominger,
``Five-branes, membranes and nonperturbative string theory,''
Nucl.\ Phys.\  {\bf B456} (1995) 130
[hep-th/9507158].


\bibitem{Harvey:1999as}
J.~A.~Harvey and G.~Moore,
``Superpotentials and membrane instantons,''
hep-th/9907026.

\bibitem{Joyce1}
D. Joyce, ``Compact Riemannian 7-manifolds with holonomy $G_2$. Part
I,'' J. Diff. Geom., {\bf 43} (1996) 291. 

\bibitem{Joyce2}
D. Joyce, ``Compact Riemannian 7-manifolds with holonomy $G_2$. Part
II,'' J. Diff. Geom., {\bf 43} (1996) 329. 

\bibitem{Bershadsky:1996sp}
M.~Bershadsky, C.~Vafa and V.~Sadov,
``D-strings on D-manifolds,''
Nucl.\ Phys.\  {\bf B463} (1996) 398
[hep-th/9510225].

\bibitem{Klemm:1996kv}
A.~Klemm and P.~Mayr,
``Strong coupling singularities and non-Abelian gauge symmetries in
$N=2$ string theory,'' 
Nucl.\ Phys.\  {\bf B469} (1996) 37
[hep-th/9601014].

\bibitem{Katz:1996ht}
S.~Katz, D.~R.~Morrison and M.~Ronen Plesser,
``Enhanced gauge symmetry in type II string theory,''
Nucl.\ Phys.\  {\bf B477} (1996) 105
[hep-th/9601108].

\bibitem{Berglund:1997uy}
P.~Berglund, S.~Katz, A.~Klemm and P.~Mayr,
``New Higgs transitions between dual N = 2 string models,''
Nucl.\ Phys.\  {\bf B483} (1997) 209
[hep-th/9605154].

\bibitem{Acharya:1998pm}
B.~S.~Acharya,
``M theory, Joyce orbifolds and super Yang-Mills,''
hep-th/9812205. 

\bibitem{Shatashvili:1994zw}
S.~L.~Shatashvili and C.~Vafa,
``Superstrings and manifold of exceptional holonomy,''
hep-th/9407025.

\bibitem{Acharya:1998rh}
B.~S.~Acharya,
``On mirror symmetry for manifolds of exceptional holonomy,''
Nucl.\ Phys.\  {\bf B524}, 269 (1998)
[hep-th/9707186].

\bibitem{Gibbons:1990er}
G.~W.~Gibbons, D.~N.~Page and C.~N.~Pope,
``Einstein metrics on $S^3$, $\mathbb{R}^3$ and $\mathbb{R}^4$ bundles,''
Commun.\ Math.\ Phys.\  {\bf 127} (1990) 529.

\bibitem{Papadopoulos:1995da}
G.~Papadopoulos and P.~K.~Townsend,
``Compactification of D = 11 supergravity on spaces of exceptional holonomy,''
Phys.\ Lett.\  {\bf B357}, 300 (1995)
[hep-th/9506150].


\bibitem{har-law}
R. Harvey and H. B. Lawson, Jr, 
``Calibrated geometries,''
Acta Math.  {\bf 148} (1982), 47.

\bibitem{Mc Lean}
R. C. McLean 
``Deformations of calibrated submanifolds,'' 
Commun. Anal. Geom. {\bf 6} (1998), 705.

\bibitem{Green:1987ck}
P.~Green and T.~H{\"u}bsch,
``Calabi-Yau manifolds as complete intersections in products of
complex projective spaces,'' 
Commun.\ Math.\ Phys.\  {\bf 109} (1987) 99.

\bibitem{Candelas:1988kf}
P.~Candelas, A.~M.~Dale, C.~A.~Lutken and R.~Schimmrigk,
``Complete intersection Calabi-Yau manifolds,''
Nucl.\ Phys.\  {\bf B298} (1988) 493.

\bibitem{Green:1989cr}
P.~S.~Green, T.~H{\"u}bsch and C.~A.~Lutken,
``All the hodge numbers for all Calabi-Yau complete intersections,''
Class.\ Quant.\ Grav.\  {\bf 6} (1989) 105.

\bibitem{Green:1987rw}
P.~Green and T.~H{\"u}bsch,
``Polynomial deformations and cohomology of Calabi-Yau manifolds,''
Commun.\ Math.\ Phys.\  {\bf 113} (1987) 505.

\bibitem{Ooguri:1996me}
H.~Ooguri and C.~Vafa,
``Summing up D-instantons,''
Phys.\ Rev.\ Lett.\  {\bf 77}, 3296 (1996)
[hep-th/9608079].

\bibitem{Greene:1996dh}
B.~R.~Greene, D.~R.~Morrison and C.~Vafa,
``A geometric realization of confinement,''
Nucl.\ Phys.\  {\bf B481}, 513 (1996)
[hep-th/9608039].



\bibitem{Sen:1996tz}
A.~Sen,
``M-Theory on $(K3 \times S^1)/\mathbb{Z}_2$,''
Phys.\ Rev.\  {\bf D53} (1996) 6725
[hep-th/9602010].

\bibitem{Kachru:2000vj}
S.~Kachru and J.~McGreevy,
``Supersymmetric three-cycles and (super)symmetry breaking,''
Phys.\ Rev.\  {\bf D61} (2000) 026001
[hep-th/9908135].

\bibitem{Joyce:1999tz}
D.~Joyce,
``On counting special Lagrangian homology 3-spheres,''
hep-th/9907013.

\bibitem{Acharya}
B.~S.~Acharya,
``On Realising N=1 Super Yang-Mills in M theory''
[hep-th/0011089].

\end{thebibliography}
\end{document}